\documentstyle[prl,aps,epsbox]{revtex}
\makeatletter
\begin{document}
\twocolumn
\newcommand{\cL}{{\cal{L}}}
\newcommand{\cH}{{\cal{H}}}
\newcommand{\cV}{{\cal{V}}}
\title{Lattice chiral symmetry with hopping interactions}
\author{Takanori Sugihara\cite{taka}}
\address{
RIKEN BNL Research Center, 
Brookhaven National Laboratory, Upton, New York 11973, USA
}
\maketitle

\begin{abstract}
We formulate Dirac fermions on a (1+1)-dimensional lattice 
based on a Hamiltonian formalism. 
The species doubling problem of the lattice fermion is 
resolved by introducing hopping interactions that 
mix left- and right-handed fermions around the momentum boundary. 
Approximate chiral symmetry is realized on the lattice. 
The deviation of the fermion propagator from 
the continuum one is small. 
\end{abstract}


In contrast with the great success of lattice gauge theory, 
lattice fermions remain a long-standing problem. 
Naive discretization causes the well-known 
species doubling problem \cite{wilson}. 
The problem originates in the fact that 
the kinetic term of the fermion is proportional to 
the first-order derivative in real space. 
This means that the Fourier transform of the kinetic term 
is proportional to the momentum. 
Since an odd function cannot be periodic and have one zero, 
the fermion propagator is forced to have additional pole(s) 
on the lattice. 
The situation does not change 
regardless of how the lattice spacing is reduced 
as long as the space-time derivative is modeled as 
a naive difference. 

Many attempts have been made to fix the doubling problem 
\cite{wilson,ks,slac,kaplan,ch,neuberger,luscher}. 
Wilson removed doublers at low energy by introducing an interaction 
that mixes left- and right-handed fermions \cite{wilson}. 
The interaction is not a result of naive discretization 
and therefore has no counterpart in continuum theory. 
However, unwanted degeneracy persist at high energy 
and chiral symmetry is explicitly broken. 
To fix these problems, 
Kaplan modified Wilson's fermion by introducing 
an extra dimension \cite{kaplan}. 
Kaplan's fermion has an approximate chiral symmetry 
if the lattice size of the extra dimension is large. 
The fermion is useful for calculating physical quantities 
related to dynamical breaking of chiral symmetry. 
(See Ref. \cite{bs}, for example.) 
However, the cost of numerical calculations 
based on Kaplan's fermion is not cheap. 
If we find a method to perform such calculations 
without the extra dimension, 
calculation time decreases largely and a deeper 
understanding of quantum field theory becomes possible. 

In addition to the doubling problem, 
the lattice fermion has another serious problem. 
The fermion propagator defined on a lattice deviates 
from the continuum one 
even if the doublers are removed with the existing 
techniques such as Kaplan's fermion \cite{ch}. 
As a result,  it may cause errors in numerical calculations. 
To remove this uncertainty, 
we need to modify the discretized propagator somehow 
so that it is close to the continuum one as far as possible. 
Such deviation of propagators becomes critical 
especially when supersymmetry is considered on a lattice 
because deviation of fermion and boson propagators 
gives wrong values for loop integrals (sums). 
For example, the zero-point energy does not cancel 
between fermions and bosons if the propagators 
deviate from the continuum ones. 
Accurate discretization of the propagators is a necessary condition 
for maintaining lattice supersymmetry. 

In general, the extra dimension can be expressed as 
hopping interactions in a lower-dimensional system. 
Kaplan's fermion is a formulation with an extra dimension, 
so there must be a corresponding Hamiltonian 
with no extra dimension. 
Also, the shape of the fermion propagator 
can be improved with hopping interactions. 
The Runge-Kutta method for differential equations 
is an example of such an improvement. 

In this paper, based on a Hamiltonian formalism, 
we introduce ultralocal hopping interactions to remove doublers 
and improve momentum dependence of fermion energy. 
(The word ``ultralocal" means that fermion hopping is 
restricted to a finite range on a real-space lattice 
\cite{bietenholz}.) 
From knowledge of the continuum theory, 
we know the correct momentum dependence of the energy. 
We start from momentum space and go back to real space 
by way of discrete Fourier transform. 
A real-space Hamiltonian is necessary to construct gauge theory. 
The method is a hybrid of the Wilson \cite{wilson}
and SLAC \cite{slac} approaches. 

\begin{figure}[h]
  \begin{center}
    \epsfile{file=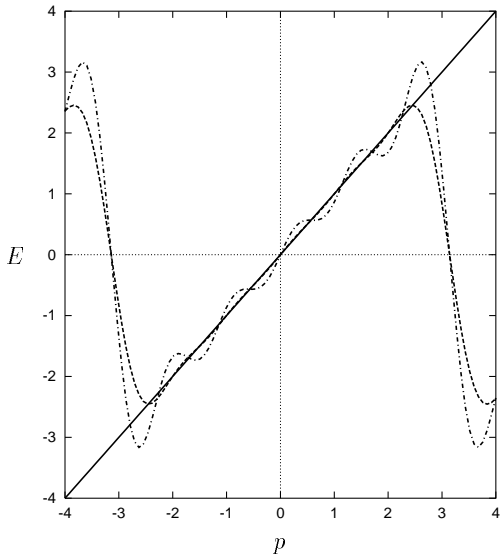,scale=0.8}
  \end{center}
  \caption{
The solid line plots the correct energy $p$ of 
one-particle states from the continuum theory. 
The dot-dashed line plots the function (\ref{sp}) for $M=5$. 
The dashed line plots the function $s(p)$ modified with 
the Lanczos factor for $M=5$. 
The oscillation of Eq. (\ref{sp}) has almost been removed. 
}
  \label{fig1}
\end{figure}

First, let us consider a free Dirac fermion 
on a (1+1)-dimensional Hamiltonian lattice. 
Time is continuous and space is discrete. 
The length of the spatial lattice is $Na$, 
where $a$ is a lattice spacing 
and the number of sites $N$ is assumed to be even. 
According to the continuum theory, 
the Dirac fermion should be described by 
the following Hamiltonian in momentum space 
\begin{equation}
 H=\frac{1}{a}\sum_{l=-N/2+1}^{N/2}
  p_l \bar{\zeta}_l \gamma^1 \zeta_l, 
 \label{h1}
\end{equation}
where $l$ is an index for momentum. 
$\zeta_l$ and $\bar{\zeta}_l$
are discrete Fourier transform 
of real-space two-component fermion operators 
\begin{equation}
 \psi_n = \pmatrix{\xi_n \cr \eta_n}
  =\frac{1}{\sqrt{N}} \sum_{l=-N/2+1}^{N/2}
  e^{i2\pi l n/N} \zeta_l, 
 \label{dft}
\end{equation}
and $\bar{\psi}_n\equiv\psi_n^\dagger\gamma^0$ 
for $n=1,2,\dots,N$, respectively. 
$\xi_n$ and $\eta_n$ satisfy 
\[
 \{\xi_m,\xi_n^\dagger\}=\{\eta_m,\eta_n^\dagger\}
 =\delta_{mn}, 
\]
and other anticommutators are zero. 
The gamma matrices are 
\[
 \gamma^0=\pmatrix{0 &  1 \cr 1 & 0}, \quad
 \gamma^1=\pmatrix{0 & -1 \cr 1 & 0}, \quad
 \gamma_5=\pmatrix{1 & 0 \cr 0 & -1}. 
\]
Periodic boundary conditions are assumed in real space, 
\[
 p_l \equiv \frac{2\pi l}{N}. 
\]
We try to create a real-space Hamiltonian 
with no doubler that reproduces Eq. (\ref{h1}). 
To find real-space representation of $p_l$, 
let us consider the following function 
\begin{equation}
 \sum_{\alpha=1}^{M}
  \frac{2(-1)^{\alpha-1}}{\alpha}
  \sin \alpha p. 
 \label{sp}
\end{equation}
In the limit $M\to\infty$, the function (\ref{sp}) goes to $p$. 
The function (\ref{sp}) necessarily has a node 
at the boundary $p=\pm\pi$ because it has a periodicity of $2\pi$ 
(see Fig. \ref{fig1}). 
The node is the cause of the doubling problem. 
The doubler remains as a singularity at the boundary 
even if the limit $M\to\infty$ is taken 
(by ``doubler" we mean unwanted energy degeneracy 
that is not contained in the continuum theory). 
Anyway, the parameter $M$ needs to be small 
for practical formulation 
because $M$ corresponds to the maximum distance 
of fermion hopping in real space. 
The limit $M\to\infty$ needs the infinite lattice. 

In addition to the doubler modes around the boundary, 
the function (\ref{sp}) has another degeneracy. 
The function oscillates around $p$ and has local minima 
if the summation is truncated with a small $M$. 
In Fourier analysis, it is called the Gibbs phenomenon, 
which  occurs 
if a function to be expanded has a singularity \cite{aw}. 
The oscillation can be removed by replacing Eq. (\ref{sp}) 
with the function 
\begin{equation}
 s(p)=\sum_{\alpha=1}^{M}
   S_\alpha \sin \alpha p, 
 \label{spl}
\end{equation}
where
\[
 S_\alpha \equiv F_\alpha
  \frac{2(-1)^{\alpha-1}}{\alpha}, 
 \quad
 F_\alpha \equiv \frac{M+1}{\pi\alpha}
   \sin\left(\frac{\pi\alpha}{M+1}\right). 
\]
$F_\alpha$ is called the Lanczos factor \cite{aw}. 
As shown in Fig.~\ref{fig1}, the factor almost removes 
the oscillation of Eq. (\ref{sp}). 
However, the doubler modes around the boundary still remain. 
We are going to remove them 
by a trick with hopping interactions. 
Let us consider the following momentum-space Hamiltonian: 
\begin{equation}
 H=\sum_{l=-N/2+1}^{N/2}
  \left(
   s_l \bar{\zeta}_l \gamma^1 \zeta_l
   +m\bar{\zeta}_l\zeta_l 
  \right), 
 \label{h2}
\end{equation}
where $s_l \equiv s(p_l)/a$ and $m$ is fermion mass. 
We write the Hamiltonian in the matrix form: 
\begin{equation}
 H=\sum_{l=-N/2+1}^{N/2} \zeta_l^\dagger
 \pmatrix{
 s_l & m \cr
 m & -s_l}\zeta_l. 
 \label{h3}
\end{equation}
We introduce an interaction $c_l$ that mixes 
left- and right-handed fermions 
\begin{equation}
 H=\sum_{l=-N/2+1}^{N/2} \zeta_l^\dagger
 \pmatrix{
 s_l & m+c_l \cr
 m+c_l & -s_l}\zeta_l, 
 \label{h4}
\end{equation}
where $c_l$ are assumed to be nonzero only for $|l|\sim N/2$. 
As shown later, $s_l$ and $c_l$ are expressed 
as ultralocal hopping interactions in real space. 
The Hamiltonian (\ref{h4}) can be diagonalized for each $l$. 
\begin{equation}
 H=\sum_{l=-N/2+1}^{N/2} \zeta_l'^\dagger
 \pmatrix{
 k_l & 0 \cr
 0 & -k_l}\zeta_l', 
 \label{h5}
\end{equation}
where $k_l\equiv \sqrt{s_l^2+(m+c_l)^2}$ are 
energies of one particle states 
and $\zeta_l'$ are transformed variables. 
Although better solutions may be found than $c_l$ shown here, 
we give precedence to simplicity over accuracy in this paper. 
We understand that our strategy is successful 
if the properties of the continuum Dirac fermion 
are approximately reproduced. 

\begin{figure}[h]
  \begin{center}
    \epsfile{file=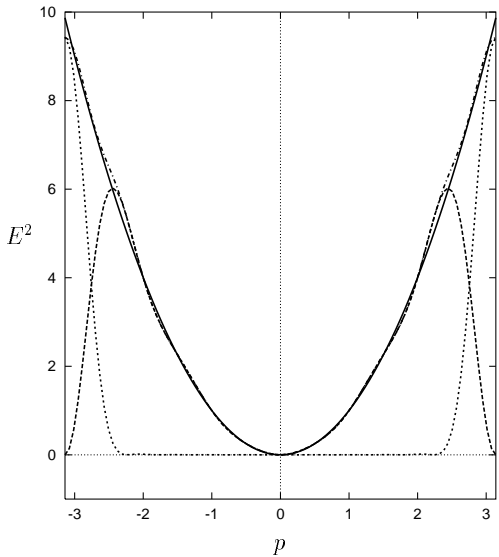,scale=0.8}
  \end{center}
  \caption{
The solid line plots the correct energy squared $p^2$. 
The dashed line plots the function $s^2(p)$ for $M=5$ 
with the Lanczos factor. 
The dotted line plots the function $c^2(p)$ for 
$M=5$, $u=130$, and $v=8.4$ with the Lanczos factor. 
The dot-dashed line plots the function $k^2(p)$. 
The doubler modes around the boundary 
$|p|=\pi$ has been removed with $c^2(p)$. 
}
\label{fig2}
\end{figure}

Consider the function 
\begin{equation}
 -u(p-\pi)^2+v=\frac{C_0}{2}
  +\lim_{M\to\infty} \sum_{\alpha=1}^M C_\alpha \cos (\alpha p), 
 \label{cfunc}
\end{equation}
where $u$ and $v$ are some positive real numbers. 
The Fourier coefficients $C_\alpha$ are 
\[
 C_0 = \frac{4v}{3\pi} \sqrt{\frac{v}{u}}, 
\]
\[
 C_\alpha = F_\alpha (-1)^\alpha \frac{4u}{\pi \alpha^2}
 \left[
  -\sqrt{\frac{v}{u}} \cos \left(\alpha\sqrt{\frac{v}{u}}\right)
  +\frac{1}{\alpha} \sin \left(\alpha\sqrt{\frac{v}{u}}\right) 
 \right]. 
\]
The function (\ref{cfunc}) has a peak at $p=\pm \pi$ and 
zeros at $p=\pm\pi\mp\sqrt{v/u}$. 
As before, the infinite $M$ cannot be realized 
on a finite lattice. 
For a finite $M$, we define the function 
\begin{equation}
 c(p) \equiv \frac{C_0}{2}+
  \sum_{\alpha=1}^M C_\alpha \cos (\alpha p). 
 \label{cpl}
\end{equation}
When $M$ is finite, $c(p)$ does not reproduce 
the function (\ref{cfunc}) correctly. 
However, it does not matter in this consideration. 
We just want to use $c(p)$ to remove the doubler modes. 
We do not need to care about the original shape 
of the function (\ref{cfunc}). 
The parameters $u$ and $v$ are adjusted so that 
the unnecessary degenerate modes around the boundary 
become non-degenerate normal modes. 
Figure \ref{fig2} shows how the doubler modes of $s(p)$ are 
removed using the function $c(p)$. 
The function $s^2(p)$ has two hemlines near the boundary 
because it has to vanish at $p=\pm \pi$. 
We want to raise the both ends 
by adding a packet function $c^2(p)$ to $s^2(p)$. 
Let us consider a function $k(p) \equiv \sqrt{s^2(p)+c^2(p)}$. 
If we choose $M=5$, $u=130$, and $v=8.4$, 
the function $k^2(p)$ agrees well with $p^2$ 
in the fundamental region $|p|\le\pi$ 
except for a small deviation around momentum $|p|\sim 2.3$. 

\begin{figure}[h]
  \begin{center}
    \epsfile{file=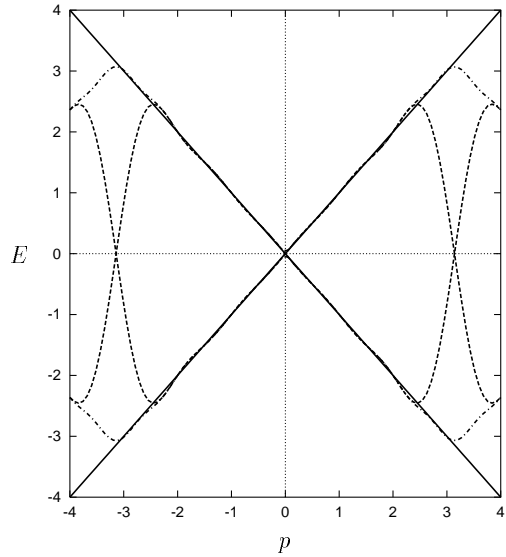,scale=0.8}
  \end{center}
  \caption{
The solid line plots the correct energy $\pm p$ 
from the continuum theory. 
The dashed line plots the function $\pm s(p)$ for $M=5$ 
with the Lanczos factor. 
The dot-dashed line plots the function $\pm k(p)$ for 
$M=5$, $u=130$, and $v=8.4$ with the Lanczos factor. 
The function $\pm k(p)$ almost agrees with $\pm p$ 
in the fundamental region $|p|\le\pi$ 
except for a small deviation around momentum $|p|\sim 2.3$. 
}
\label{fig3}
\end{figure}

In Fig. \ref{fig3}, the functions $\pm p$, $\pm s(p)$, 
and $\pm k(p)$ are compared. 
The function $\pm k(p)$ corresponds to one-particle 
states given by the Hamiltonian (\ref{h4}) with $m=0$. 
The function $\pm k(p)$ agrees well with the correct 
energy $\pm p$ from the continuum theory 
in the fundamental region $|p|\le\pi$. 
The Hamiltonian (\ref{h4}) approximately reproduces 
the continuum theory for the (1+1)-dimensional Dirac spinor 
without doubler modes 
if we identify $c_l = c(p_l)/a$ and $k_l=k(p_l)/a$.

In the new basis that diagonalizes Eq. (\ref{h4}) with $m=0$, 
$\gamma_5$ is transformed into 
\begin{equation}
 \gamma_5' = 
  \frac{s_l+k_l}{k_l^2+s_l k_l}
  \pmatrix{s_l & -c_l \cr -c_l & -s_l} 
 \label{g5}
\end{equation}
for $l>0$ and 
\begin{equation}
 \gamma_5' = 
  \frac{-s_l+k_l}{k_l^2-s_l k_l}
  \pmatrix{s_l & c_l \cr c_l & -s_l} 
 \label{g5p}
\end{equation}
for $l<0$. For $l=0$, the Hamiltonian (\ref{h4}) is diagonal 
with degenerate zero energy, so $\gamma_5$ is not transformed. 

\begin{figure}[h]
  \begin{center}
    \epsfile{file=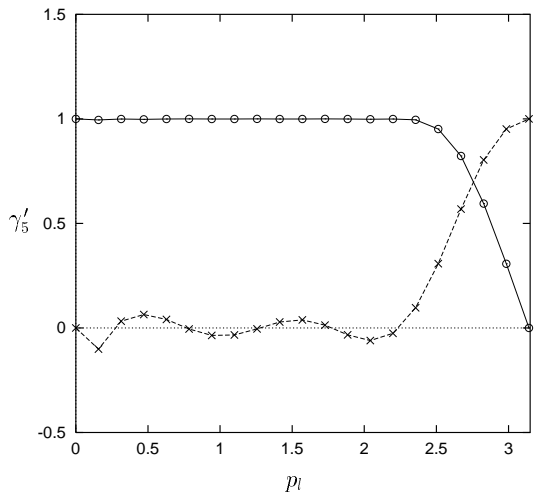,scale=0.8}
  \end{center}
  \caption{
The matrix elements of Eq. (\ref{g5}) are plotted as functions 
of the momentum $p_l$ for $a=1$, $M=5$, $u=130$, and $v=8.4$ 
with the Lanczos factor. 
For a lattice size $N=40$, the functions are plotted 
in the right-half plane of momentum space $p_l\ge 0$. 
The circles are plots of the diagonal (1,1) element 
of Eq. (\ref{g5}). The crosses are plots of minus 
of the off-diagonal (1,2) element of Eq. (\ref{g5}). 
The solid and dashed lines are plotted to guide the eyes. 
}
\label{fig4}
\end{figure}

Figure \ref{fig4} shows the diagonal and off-diagonal 
matrix elements of the transformed $\gamma_5'$. 
The diagonal (1,1) element of Eq. (\ref{g5}) is 
almost unity at low and intermediate energy $p_l<2.3$ 
and deviates from unity at $p_l>2.3$. 
The off-diagonal (1,2) element of Eq. (\ref{g5}) oscillates 
around zero at $p_l<2.3$ and becomes unity at $p_l=\pi$. 
At low energy, the deviation of the off-diagonal elements 
from zero is not large and becomes smaller 
as the parameter $M$ increases. 
At $p_l<2.3$, the transformed left- and right-handed fermions 
have approximately the correct chiral charges $1$ and $-1$, 
respectively. 
The low-energy Hamiltonian (\ref{h4}) has approximate 
chiral symmetry because the commutation relation between 
the Hamiltonian (\ref{h4}) and chiral charge defined 
with $\gamma_5'$ is almost zero for small $l$. 
The errors associated with chiral symmetry can be improved 
in a systematic way by increasing $M$. 
The value used here for the parameter $M$ is sufficiently small 
and does not deny application of the model to 
actual numerical analysis with a computer. 

To obtain the real-space Hamiltonian for the Dirac spinor, 
we substitute the discrete Fourier transform (\ref{dft}) 
into Eq. (\ref{h4}) 
\begin{eqnarray}
 &&H= \sum_{n=1}^N
 \bigg\{
 \frac{1}{2a} \sum_{\alpha=1}^M
 \Big[
  iS_\alpha ( \bar{\psi}_{n+\alpha} \gamma^1 \psi_n
    -\bar{\psi}_n \gamma^1 \psi_{n+\alpha} )
\nonumber
\\
  &&
  +C_\alpha ( \bar{\psi}_{n+\alpha}\psi_n
    +\bar{\psi}_n\psi_{n+\alpha} ) 
 \Big]
 +\left(m+\frac{C_0}{2a}\right)\bar{\psi}_n\psi_n
 \bigg\}. 
 \label{h6}
\end{eqnarray}
As usual, gauge symmetry can be implemented by 
inserting exponentiated gauge fields 
between the hopping fermions in Eq. (\ref{h6}). 
The continuum limit $a\to 0$ is taken with 
the parameter $M$ fixed. 
The method can be extended to higher-dimensional cases 
by taking care of the doubler of each direction 
in the same way. 

For Monte Carlo analysis, 
it is useful to rewrite our Hamiltonian into 
a Euclidean formulation. 
The Euclidean action for the Dirac spinor is given by 
\begin{eqnarray}
 &&S_{\rm E}=\sum_n 
 \bigg\{ \frac{1}{2a}\sum_{\alpha=1}^M \sum_{\mu=1}^2
 \bigg[S_\alpha
  \big(
    \bar{\psi}_n \gamma_\mu \psi_{n+\alpha\hat{\mu}}
   -\bar{\psi}_{n+\alpha\hat{\mu}} \gamma_\mu \psi_n
  \big)
 \nonumber
 \\
 &&
 +C_\alpha
  \big(
   \bar{\psi}_n \psi_{n+\alpha\hat{\mu}} + 
   \bar{\psi}_{n+\alpha\hat{\mu}}\psi_n
  \big)
 \bigg]
 +\left(m+\frac{C_0}{a}\right)\bar{\psi}_n \psi_n
 \bigg\}, 
 \label{ea}
\end{eqnarray}
where $n$ indicates a site on a two dimensional Euclidean 
lattice and $\hat{\mu}$ is a unit one-site vector 
in the $\mu$ direction. 
The Euclidean gamma matrices satisfy 
$\gamma_\mu^\dagger=\gamma_\mu$ and 
$\{\gamma_\mu,\gamma_\nu\}=2 \delta_{\mu\nu}$ 
and $\bar{\psi}\equiv\psi^\dagger\gamma_2$. 
The definition of a lattice gauge theory based on the action 
(\ref{ea}) is ambiguous. 
An exponentiated gauge field can take 
any path between two ends. 
The most natural choice for gauge field 
is the shortest path that links the two ends 
because fermion hopping is parallel to 
one of the directions $\hat{\mu}$. 

In this paper, we have constructed doubler-free 
Hamiltonian and Euclidean action for Dirac fermions 
on a (1+1)-dimensional lattice. 
To realize approximate chiral symmetry at low energy, 
explicit breaking of chiral symmetry has been compressed to 
around the momentum boundary. 
In future works, it should be precisely checked 
if insertion of gauge interactions affects chiral properties. 

I am indebted to Michael Creutz for valuable discussions. 
I also thank Jun-ichi Noaki for informative discussions. 
This research was supported in part by RIKEN.

\end{document}